\documentclass[12pt]{article}
\pdfoutput=1
\usepackage{color}
\usepackage{amsmath, amssymb}
\usepackage{epsfig, palatino}
\usepackage{pstricks,pst-node,pst-tree}
\usepackage{epic}

\usepackage{mathrsfs}

\usepackage{ae} % or {zefonts}0
\usepackage[T1]{fontenc}
\usepackage[ansinew]{inputenc}
\usepackage{amsmath}
\usepackage{amssymb}
\usepackage{graphicx}
\usepackage{color}
\definecolor{darkblue}{cmyk}{0.9,0.9,0,0}
\usepackage[colorlinks=true,linkcolor=darkblue,citecolor=darkblue,urlcolor=darkblue]{hyperref}
\newcommand{\beq}{\begin{equation}}
\newcommand{\eeq}{\end{equation}}
\newcommand\beqa{\begin{eqnarray}}
\newcommand\eeqa{\end{eqnarray}}
\newcommand\bea{\begin{array}}
\newcommand\eea{\end{array}}

\def\XXint#1#2#3{{\setbox0=\hbox{$#1{#2#3}{\int}$}
\vcenter{\hbox{$#2#3$}}\kern-.5\wd0}}

\newcommand{\nn}{\nonumber}

\newcommand{\COMMENT}[1]{}

\newcommand{\neqa}{\nonumber\end{eqnarray}}
\newcommand{\la}[1]{\label{#1}}

\newcommand{\<}{{\langle}}
\renewcommand{\>}{{\rangle}}

\newcommand{\cA}{{\cal A}}

\newcommand{\re}{\relax{\rm I\kern-.18em R}}

\def\su2{{SU(2)}}

\def\[{\left[}
\def\]{\right]}

\def\({\left(}
\def\){\right)}
\def\[{\left[}
\def\]{\right]}

\def\<{\langle}
\def\>{\rangle}

\def\i2{\frac{i}{2}}

\def\cF{{\cal F}}
\def\cC{{\cal C}}

\usepackage{varioref}
\usepackage{makeidx}
\makeindex
\usepackage[english]{babel}

\begin{document}

%%%%%%%%%%%%%%%%%%%%%%%%%%%%%%%%%%%%%
%%%%%%%%%%%%%%%%%%%%%%%%%%%%%%%%%%%%%
%%%%%%%%%%%%%%%%%%%%%%%%%%%%%%%%%%%%%
%

\thispagestyle{empty}
%\begin{flushright}\footnotesize
%\texttt{arxiv:yymm.nnnn}\\
%\texttt{AEI-2009-078}\\
%\vspace{1.7cm}
%\end{flushright}

\renewcommand{\thefootnote}{\fnsymbol{footnote}}
\setcounter{footnote}{0}
\setcounter{figure}{0}
\begin{center}
$$$$
{\Large\textbf{\mathversion{bold}
Multichannel Conformal Blocks\\ for Polygon Wilson Loops}\par}

\vspace{1.0cm}
\textrm{Amit Sever, Pedro Vieira}
%\\ \vspace{1.2cm}
\footnotesize{

\textit{
Perimeter Institute for Theoretical Physics\\ Waterloo,
Ontario N2J 2W9, Canada} \\
\texttt{}
\vspace{3mm}
}

\par\vspace{1.5cm}

\textbf{Abstract}\vspace{2mm}
\end{center}
%
%\noindent
%%\small
%
%We consider the Operator Product Expansion for Null Polygon Wilson loops with more than six edges. These objects are the analogues of higher point correlation functions of local operators. To study them we introduce the notion of higher point conformal blocks for Polygon Wilson Loops which we denote as Multichannel Blocks. As an application of the formalism we decompose the one loop Heptagon Wilson loop and predict the value of its two loop OPE discontinuities. At the functional level, the OPE discontinuities are roughly half of the full result. Using symbols they suffice to predict the full two loop result. We also present several new predictions for parts of the Heptagon result at any loop order. \\
We introduce the notion of Multichannel Conformal Blocks relevant for the Operator Product Expansion for Null Polygon Wilson loops with more than six edges. As an application of these, we decompose the one loop heptagon Wilson loop and predict the value of its two loop OPE discontinuities. At the functional level, the OPE discontinuities are roughly half of the full result. Using symbols they suffice to predict the full two loop result. We also present several new predictions for  the heptagon result at any loop order.

\vspace*{\fill}

\setcounter{page}{1}
\renewcommand{\thefootnote}{\arabic{footnote}}
\setcounter{footnote}{0}

\newpage
%\tableofcontents
\section{Introduction and Review}
Null Polygon Wilson loops in $\mathcal{N}=4$ SYM are extremely interesting objects. One of the reasons is because they yield all planar scattering amplitudes of the theory \cite{AmplitudeWilson}.\footnote{The usual bosonic loops, which are the ones considered in this note, lead to Maximally Helicity Violating amplitudes. A proper supersymmetrization of bosonic loop leads to all scattering amplitudes.} In \cite{OPEpaper} an operator product expansion (OPE) for Null Polygon Wilson loops (NPWL) was proposed; it is the analogue of the usual operator product expansion for local operators in a conformal field theory. 
One of the main ingredients in the study of the latter are functions which package together the propagation of a conformal primary and all its descendants in a conformal field theory. These are known as \textit{conformal blocks} \cite{blocks1,blocks2}. The main focus of this brief note is on the OPE for NPWL with more than six edges and in particular on their corresponding conformal blocks. They were first used in this setup in \cite{Hexagonpaper} in the re-derivation of the two loop hexagon Wilson loop \cite{DelDuca,Gon}.

\subsection*{OPE review for the simplest kinematics}
Let us recall how to interpret simple known results for NPWL from the OPE picture. For more details see \cite{OPEpaper,Hexagonpaper,bootstraping}. We start by discussing some kinematics before turning to the dynamics.
The simplest NPWL are polygons which lie in a $\mathbb{R}^{1,1}$ plane \cite{AMapril}.  The cusps of such polygons are given by light-cone coordinates $\(x_i^+ , x_i^-\)$. The conformal symmetry group 
 acting on such polygons reduces to $SL(2)_+ \times SL(2)_-$ generated by
\beq
L^{(i)}_{-1} = - \frac{\partial}{\partial {x_i^+}} \,, \qquad L^{(i)}_{0} = - {x_i^+} \frac{\partial}{\partial {x_i^+}}  \,,\qquad L^{(i)}_{1} = -\({x_i^+}\)^2  \frac{\partial}{\partial {x_i^+}}
\eeq
with similar generator $\bar L_a^{(i)}$ acting on $x_i^-$.
Conformal cross-ratios factorize into a product of {left} and {right} cross-ratios $\chi_{ijkl}^+ \equiv  \frac{(x_i^+-x_j^+)(x_k^+-x_l^+)}{(x_i^+-x_l^+)(x_k^+-x_j^+)}$ and $\chi_{ijkl}^- \equiv  \frac{(x_i^--x_j^-)(x_k^--x_l^-)}{(x_i^--x_l^-)(x_k^--x_j^-)}
$
which are invariant under $SL(2)_+ \times SL(2)_-$ transformations acting on \textit{all cusps}. Such transformations are generated by 
\beq
L_a= \sum_{i=1}^n L_a^{(i)} \qquad \text{and} \qquad \bar L_a= \sum_{i=1}^n \bar L_a^{(i)}
\eeq
where $2n$ is the number of cusps. 
\begin{figure}[t]
\centering
\def\svgwidth{18cm}
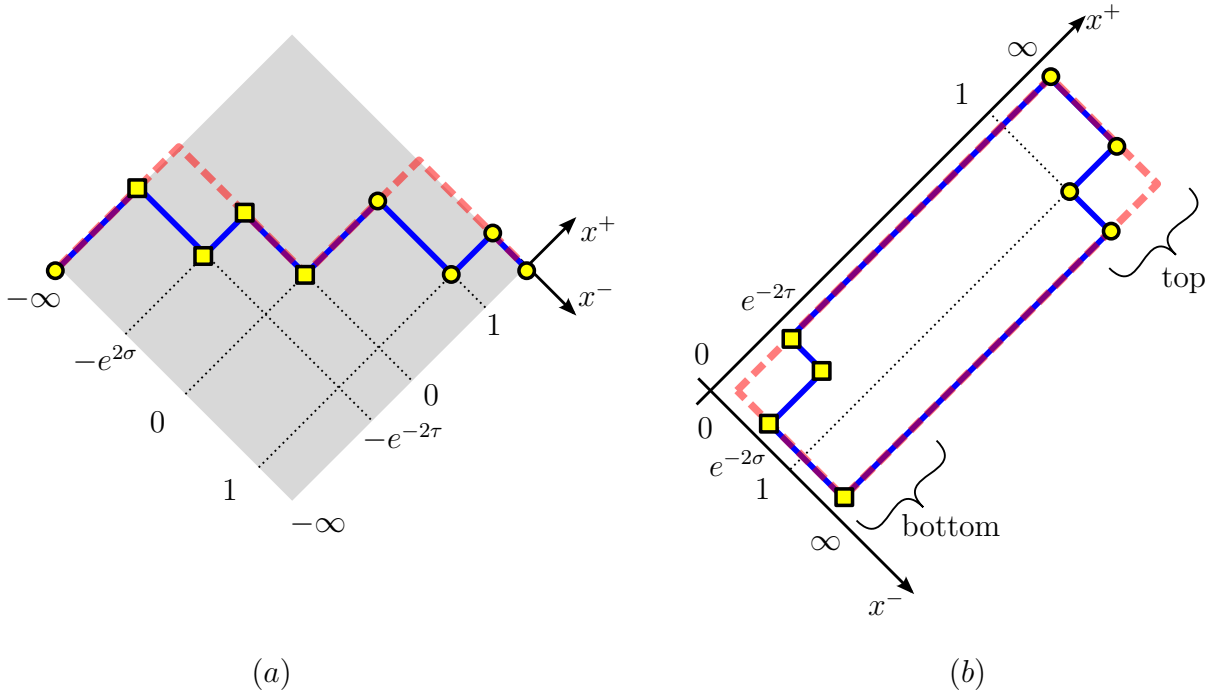
\caption{The octagon NPWL in $\mathbb{R}^{1,1}$ kinematics. (a) The Euclidean picture where all points are spacelike or null separated. This is the kinematical regime considered in this paper. (b) A more physically intuitive Lorentzian picture. The results for the two pictures are related by analytic continuation.
}\label{octagonfig}
\end{figure}
The first non-trivial polygon occurs when we have four $x_i^+$ and four $x_i^-$, that is for $n=4$. This is the octagon represented in  figure \ref{octagonfig}. It contains only two independent cross ratios which can be chosen as $\chi^+ =e^{-2\tau}$ and $\chi^- = e^{2\sigma}$.
One configuration of $x_i^{\pm}$ leading to these cross ratios is depicted in figure \ref{octagonfig}a. The variables $\tau$ and $\sigma$ have one further nice geometrical meaning as we now explain. Given a NPWL we can define a null reference square represented by the dashed lines in figure \ref{octagonfig}. A choice of square automatically splits the polygon cusps into two groups: the \textit{top} and the \textit{bottom} as represented in the figure.\footnote{In the usual OPE for local operators we pick a conformal frame which consists of two points, conventionally chosen to be zero and infinity. The conformal generators used for the OPE (dimension and spin) are those that leave the conformal frame invariant. The reference square is the conformal frame in the Wilson loops picture.}
There is a $\mathbb{R}_+ \times \mathbb{R}_- \subset SL(2)_+ \times  SL(2)_- $ residual conformal symmetry which leaves the reference square invariant. For our choice of points, the residual symmetry $\mathbb{R}_+$ ($\mathbb{R}_-$) is generated by $L_0$ ($\bar L_0$). Consider the action of these symmetries on \textit{one of the groups}, say the bottom one. That is, consider  
\beq
\mathcal{L}_0= \sum_{i\in\text{bottom}} L_0^{(i)} =-2\partial_{\tau} \qquad  \text{and} \qquad \bar{ \mathcal{L}}_0= \sum_{i\in\text{bottom}} \bar L_0^{(i)}=2 \partial_{\sigma}  \,.
\eeq
We see that the action of the residual symmetry is trivial: it amounts to translations of $\tau$ and $\sigma$. 
An equivalent way of rephrasing this is the following. We start with any octagon, for example the one with $\tau=\sigma=0$. Then we generate a family of polygons by acting on the bottom cusps with the symmetries of the reference square. For the octagon in $\mathbb{R}^{1,1}$ we have two symmetries and two independent conformal-cross ratios. Hence we describe all possible octagons in this way.

\begin{figure}[t]
\centering
\def\svgwidth{10cm}
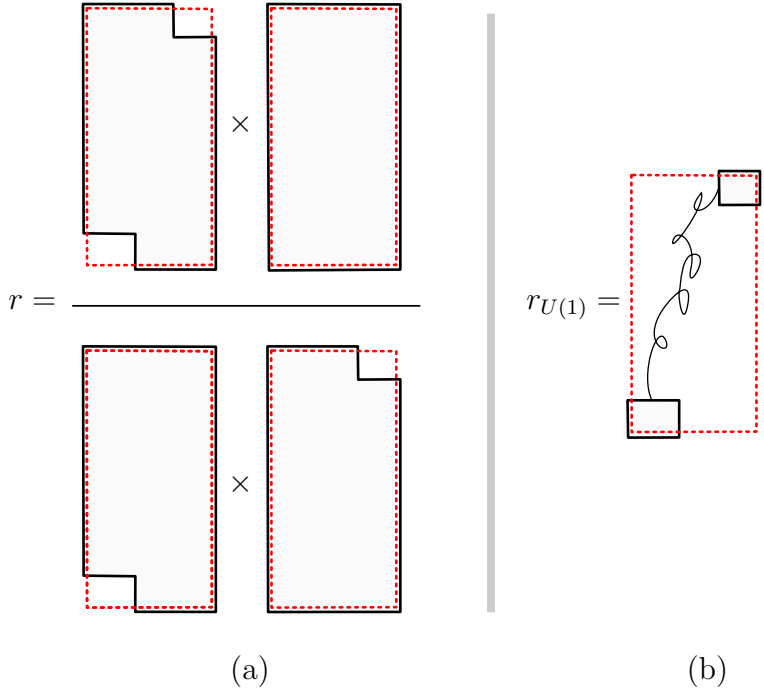
\caption{Octagon ratio $r$. At one loop this ratio is computed by the disconnected correlator of two rectangle Wilson loops. In the OPE picture excitations are produced in the bottom rectangle and absorbed by the top rectangle; in between they propagate \textit{freely} since at this loop order there are no interactions with the flux tube. That is, the gluon ``doesn't know" about the reference square. 
}\label{roctagon}
\end{figure}

Having discussed the kinematics we  move to the discussion of the Wilson loop expectation value. We will continue to base our discussion on the simplest example, the octagon in $\mathbb{R}^{1,1}$.
Wilson loops with cusps have well understood UV divergences coming from virtual gluons exchange between neighboring edges. A regularization of these divergences breaks conformal invariance in a controlled way \cite{Drummond:2007aua}. 
Since they are well understood it is simple to subtract them. What remains is a well defined conformal invariant function dubbed the \textit{remainder function}. An alternative observable is constructed by taking a ratio of NPWLs  such that the divergences cancel out and one remains with a conformal invariant finite function named $r$ in \cite{OPEpaper}. It is related to to the remainder function in a trivial way \cite{OPEpaper,bootstraping}. For the octagon, that ratio of Wilson loops is 
\beq
r^{\text{octagon}}=\log \frac{\<W^{\text{octagon}}\>\<W^{\text{reference square}}\>}{\<W^{\text{top hexagon}}\>\<W^{\text{bottom hexagon}}\>} \,,
\eeq
see figure \ref{roctagon}a. At one loop this quantity is given by the correlation function of the two disconnected Wilson loops as depicted in figure \ref{roctagon}b. If two opposite cusps of the reference square are chosen to coincide with two cusps of the octagon (as in the figure), then all divergences exactly cancel.\footnote{In other words, the total dual conformal anomaly \cite{Drummond:2007aua} is zero for this quantity.}
We have \cite{bootstraping}
\beq
r^{\text{octagon}}_{\text{1 loop}}(\tau,\sigma)= -\frac{g^2}{2} \log(1+e^{-2\tau})\log(1+e^{-2\sigma}) \,, \la{roneloop}
\eeq
which is indeed a conformal invariant function of the two cross-ratios of the octagon. It is designed such that only excitations that propagate from the bottom to the top survive.   

Let us try to derive this expression from the OPE picture, that is without doing any Feynman diagram.  
We want to think of excitations being created at the bottom of the Wilson loop in figure \ref{octagonfig}b, propagating in the middle region and being absorbed by the top part of the Wilson loop. 
In the propagating region the excitations interact with the left and right null edges of the reference square. Those two edges create a flux tube between them hence  the excitations we have in mind are flux tube perturbations \cite{OPEpaper}.  The flux tube excitations will have two important quantum numbers: energy (or twist) $E$ conjugated to the $\tau$ translation and momentum $p$ conjugated to the $\sigma$ translation. 
To measure what is flowing we construct the Casimir of the $SL(2)_+$ conformal group,
\beq
\mathcal{C} \equiv 2\mathcal{L}_{0}^2   -\mathcal{L}_1 \mathcal{L}_{-1} - \mathcal{L}_{-1} \mathcal{L}_{1} 
\eeq
where
\beq
\mathcal{L}_a= \sum_{i\in\text{bottom}} L_a^{(i)}
\eeq
Note the following very important point.\footnote{What we are doing mimics very closely the treatment of conformal blocks of Dolan and Osborn for correlation function of local operators  \cite{blocks2}.}
The generators $\mathcal{L}_a$ act on the bottom part of the polygon only. This is what we want to do since we want to measure what was created there and is propagating in the direction of the top region. In our parametrization we have
\beq
\mathcal{C} = \frac{e^{-2\tau}+1}{2} \, \frac{d^2}{d \tau^2} +\frac{d}{d \tau}   \la{casimiroctagon}
\eeq
%Provided the vacuum, represented by the flux, is $SL(2)$ invariant then 
The action of $\cC$ on a primary excitation of twist $E$ and its conformal descendants yields the casimir $2E(E-2)$.\footnote{An important assumption is that the vacuum, represented by the flux, is $SL(2)$ invariant; this is true at the first loop orders but eventually it breaks down \cite{Benjamin,bootstraping}. For the purpose of the current paper we do not need to worry about the higher loop breakdown. When studying higher loops it might be important to understand more precisely how this is broken.} 
 In the two dimensional kinematics, the excitations are just the primary $F_{+-}$ and its descendents $(D_-)^{k-1} F_{+-}$. These are the excitations which are generated by deformations of the loop in $\mathbb{R}^{1,1}$. The twist of the primary $F_{+-}$ is $E=2$. % and at one loop, the vacuum is indeed $SL(2)$ invariant.\footnote{At higher loops, the flux breaks the $SL(2)$ symmetry \cite{OPEpaper}.}  
 Hence we should find $\mathcal{C}\, r^{\text{octagon}}_{\text{1 loop}}(\tau,\sigma) = 0$ which implies that $ r^{\text{octagon}}_{\text{1 loop}}(\tau,\sigma) = c_1(\sigma) \log(1+e^{-2\tau})+c_2(\sigma)$. The result should vanish in the OPE limit $\tau\to \infty$ and it should be $\sigma \leftrightarrow \tau$ symmetric (this is just parity, see figure \ref{octagonfig}). Hence $c_2(\sigma)=0$ and $c_1(\sigma) = c \log(1+e^{-2\sigma})$. Up to the undetermined constant $c$ related to the strength of the coupling constant, we just derived the octagon one loop result (\ref{roneloop}) from conformal symmetry. In this example, we see that 
\beq
\mathcal{F}(\tau)\equiv  \log(1+e^{-2\tau}) =\sum_k \frac{(-1)^k}{k} e^{-2k\tau}   \la{octblock}
\eeq
describes the propagation of the primary and all its descendents. 
In the usual OPE language we would rephrase this by saying that $\mathcal{F}(\tau)$ 
is a $SL(2)_+$ conformal block describing the propagation of a primary of twist $E=2$ and all its descendents. 
It is fixed by the differential equation $\mathcal{C} \mathcal{F}(\tau)=2E(E-2)\mathcal{F}(\tau)$ with the appropriate boundary conditions. 

What about higher loops? 
To think about what could change it is instructive to re-write (\ref{roneloop}) in the following inspiring form \cite{OPEpaper}, 
\beq
r_{\text{1 loop}}^{\text{octagon}}(\tau,\sigma)=\sum_k \int dp \,C_k^{(1)} (p) \,e^{-i p \sigma - E^{(0)}_k(p) \tau}
\eeq
where $E^{(0)}_k(p)=2k$ is the free twist of $F_{+-}$ and its descendents and the form factor $C_k^{(1)}(p)$ is interpreted as the probability amplitude for creating the excitation with energy $E_k^{(0)}$ at the bottom and absorbing it at the top. It can be computed by Fourier transforming the one loop result. The momentum $p$ is the quantum number conjugate to the translations in $\sigma$ which is a symmetry of the reference square. 
There are three types of higher loop corrections: 
(a) We can start having more than a single particle propagating,\footnote{We have gapped excitations on the infinite line and therefore can talk about particles.} (b) the form factors can get corrections and finally (c) the classical energy of the excitations will acquire an anomalous contribution from the interaction with the flux tube. The latter has quite a distinctive feature in perturbation theory: it appears in the exponent multiplying $\tau$. The expansion of the exponential in the 't Hooft coupling $g^2$, leads to a term linear in $\tau$, i.e. we have
\beq
r^{\text{octagon}}_{\text{2 loops}}(\tau,\sigma) = \underbrace{ \sum_{k} C_k^{(2)}(\sigma) e^{-k \tau} }_{\text{from contributions (a) and (b)}}+ \tau \underbrace{ \sum_{k} D_k^{(2)}(\sigma) e^{-k \tau} }_{\text{from contribution (c)}}
\eeq
The anomalous dimension of the excitations $D_-^{k-1} F_{+-}$ reads \cite{bootstraping,Benjamin}
\beq
E^{(1)}_k(p) = 2g^2 \[\psi(1+i p/2)+\psi(1-i p/2)-2\psi(1)\] \equiv g^2 \gamma_2(p)
\eeq
In particular, it is independent of $k$. This is to be expected since these excitations are descendents, related to $F_{+-}$ by the $SL(2)$ symmetry that is not broken at this loop order.\footnote{At two loops however, the SL(2) symmetry is broken and the two loop energy $E^{(2)}_k$ starts to depends on $k$.}  Hence we can easily compute the contribution linear in $\tau$ of the two loop result,\footnote{This agrees neatly with the known result \cite{DelDucaOct} as shown in \cite{bootstraping}.}
\beq
\left.r_{\text{2 loop}}^{\text{octagon}}(\tau,\sigma) \right|_{\text{linear in }\tau}= -\sum_k \int dp \,C_k^{(1)} (p) \,\gamma_2(p)\,e^{-i p \sigma - 2k  \tau} \la{2loopoct}
\eeq
Similarly, at $l$ loops the OPE expansion indicates that
\beqa
r_{\text{l loops}}^{\text{octagon}} &=&  \sum_{k} C_k^{(l)}(\sigma) e^{-k \tau} + \tau  \sum_k D_k^{(l)}(\sigma) e^{-k \tau} + \dots \nn \\
&+& \frac{(-1)^{l-1}\,\tau^{l-1} }{(l-1)!}\sum_k  \int dp \,C_k^{(1)} (p) \,\gamma_2(p)^{l-1}\,e^{-i p \sigma - 2k  \tau}\la{Nloopoct}
\eeqa
This provides an infinite amount of predictions for the octagon Wilson loop at arbitrary loop order. At two loops this is enough to constrain the result completely \cite{bootstraping,Hexagonpaper}.\footnote{Which agrees with earlier predictions \cite{DelDucaOct,Heslop:2010kq}.} At higher loops it provides important constraints but it does not seem to be enough, one needs to consider also other $\tau^{m<l-1}$ contributions coming from corrections to form factors, multi-particles, higher loop corrections to the anomalous dimensions etc. 

\subsection*{New ingredients for general kinematics}

\begin{figure}[t]
\centering
\def\svgwidth{6cm}
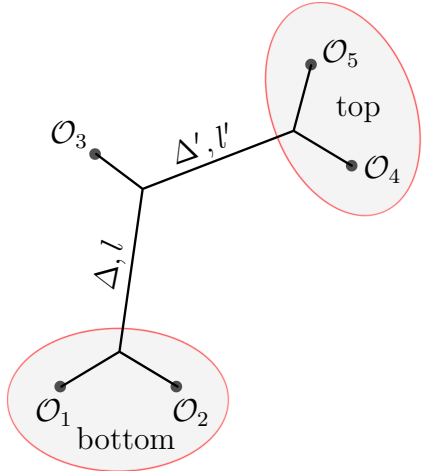
\caption{A five point function of local operators. It is the correlation function analog of the heptagon NPWL considered in this paper. For the five point function, the generalized conformal blocks describe the propagation of two primaries and their conformal descendants in two channels simultaneously. In the figure, the two primaries are parametrize by their dimension ($\Delta$) and spin ($l$).}\label{fivepoint}
\end{figure}
Let us summarize the discussion so far and point out the new ingredients that come into play when considering more general polygons (they are the main focus of this short note)
\begin{itemize}
\item For the octagon in $\mathbb{R}^{1,1}$ we have $2$ independent cross-ratios. For a polygon in $\mathbb{R}^{1,3}$ with $n$ edges we have $3n-15$ independent cross-ratios. 
\item For the octagon in $\mathbb{R}^{1,1}$ we could parametrize the $2$ cross-ratios by the symmetries of the reference square. The main advantage of this parametrization is that it allows us to make use of the OPE physical picture in a straightforward way. 
For the hexagon in $\mathbb{R}^{1,3}$ we have three cross-ratios and this is still possible. For this we use the two symmetries that we used before plus an extra symmetry of the reference square which we did not make use so far: the $SO(2)$ rotations in the two directions transverse to the plane where the square is. The square has no more symmetries. Hence we can not parametrize the family of all heptagons (or any other $\mathbb{R}^{1,3}$ polygons with more than six edges) in such simple geometrical way. In this paper we propose a parametrization of higher $n$ polygons using more than one reference square. This will still allow us to make use of the OPE in a very efficient way. For example we will predict what the analogue of (\ref{2loopoct}) and (\ref{Nloopoct}) is for the heptagon Wilson loop in $\mathbb{R}^{1,3}$.
\item For polygons in ${\mathbb R}^{1,1}$ the relevant primaries which we should consider at one and two loops are $F_{+-}$ excitations. For polygons in ${\mathbb R}^{1,3}$ there are two infinite towers of primaries which we need to consider \cite{OPEpaper,Hexagonpaper} (they are labelled by the transverse $SO(2)$ charge which played no role for the two dimensional kinematics). Hence, we will have to deal with an infinite tower of conformal blocks instead of the single block (\ref{octblock}) considered above. For the hexagon polygon this was done in \cite{Hexagonpaper}.

\item Polygons in  ${\mathbb R}^{1,3}$ with more than six edges are the analogue of correlation function for local operators with more than four points. The kind of conformal blocks that we need to introduce depends on more variables and capture the propagation in more than one channel simultaneously. For example, for the five point correlation function as well as for the heptagon NPWL, these blocks describes the propagation of two primaries and their conformal descendants in two different channel at the same time (see figure \ref{fivepoint}). The computation of these blocks on the heptagon is our main result. It will allow us to decompose the one loop result in an OPE friendly way and therefore it will allow us to make an infinite number of predictions for the heptagon at any loop order. 
\end{itemize}

In this paper, we explain how to bootstrap polygons with more than six edges using the OPE program of \cite{OPEpaper}. The basic novelty of the present approach is the consideration of more than one OPE channel simultaneously. To illustrate the power of our approach, we apply it to the heptagon NPWL. We compute  its multi-channel conformal blocks and use these to decompose the one loop result in an OPE friendly way.  It will allow us to compute the heptagon OPE discontinuities and make an infinite number of predictions for the heptagon at any loop order. The paper is organized as follows. Section \ref{sec2} is the bulk of this short note where we derive the heptagon multi-channel conformal blocks, decompose the one loop result and use it to compute the OPE discontinuity. In section \ref{sec3} we discuss the generalization to NPWL with more edges. The appendix include some technical details for the heptagon.

\section{The heptagon Wilson loop}\la{sec2}

\begin{figure}[t]
\centering
\def\svgwidth{15cm}
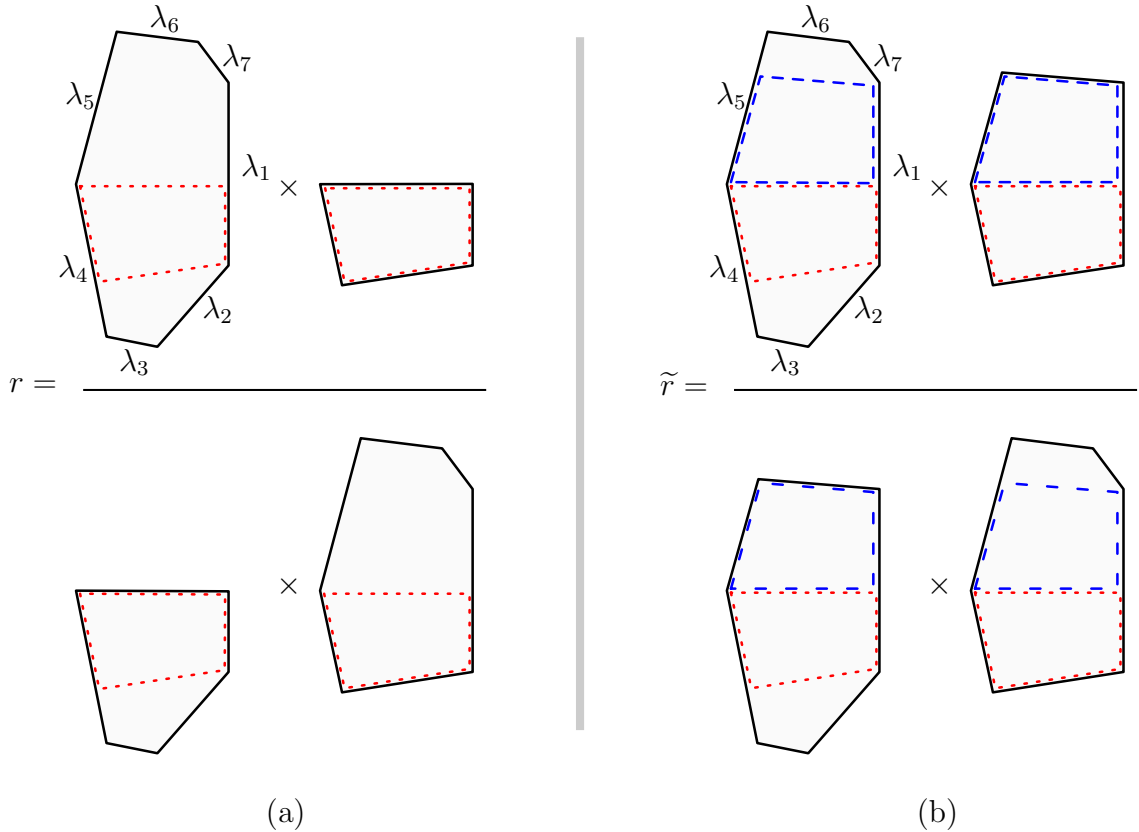
\caption{(a) Finite, conformal invariant ratio of Wilson loops using a single reference square. (b) Finite, conformal invariant ratio of Wilson loops using two reference squares. }\label{TopBottom}
\end{figure}
In the two dimensional kinematics the first NPWL with non-trivial cross-ratios is the octagon. In four dimensional kinematics, the first non-trivial polygons is the hexagon. These are the analogues of the four point function in the context of correlation functions of local operators. This paper is about polygons with more edges which are the analogue of higher point correlation functions.

In this section we focus on the heptagon in $\mathbb{R}^{1,3}$ which is the Wilson loop cousin of the five point correlation function of local operators. It contains all the new ingredients involved in the generalization from six edges to higher and hence serves well the purpose of explaining the general method. Our main result of this section are the decomposition of the heptagon one loop result in multi-channel conformal blocks presented in (\ref{form}-\ref{theC}) and the prediction of its two loops OPE discontinuity presented in (\ref{twoloops}).
The generalization to higher polygons is discussed in the next section. 

The first step in bootstrapping the heptagon is to choose an OPE channel and a corresponding conformal frame. That is, to choose a reference square. For the heptagon, all such choices are related by cyclic permutations. One then constructs the corresponding ratio of polygons (see figure \ref{TopBottom}.a)
\beq\la{smallr}
r=\log {\<W^{\text{heptagon}}\>\<W^{\text{red dotted square}}\>\over\<W^{\text{top hexagon}}\>\<W^{\text{bot pentagon}}\>} \,.
\eeq
This is a finite conformal invariant observable. The function $r$ is associated with a specific channel and therefore, contrary to the remainder function $R$, it is not a cyclic invariant function. The relation between the two is simply
%
%It is related to the remainder function $R_{\text{heptagon}}$ in a simple way
\beq
R_{\text{heptagon}}=r-r_{\widetilde U(1)}-R_\text{top hexagon}-R_\text{bot pentagon}
\eeq
where $r_{\widetilde U(1)}$ is the  ratio (\ref{smallr}) computed in a $U(1)$ theory and dressed by the cusp anomalous dimension \cite{OPEpaper}. In other words, $r_{\widetilde U(1)}$ is the  ratio (\ref{smallr}) obtained by replacing the Wilson loop expectation values by the BDS result \cite{BDS}. An advantage of $r$ over $R$ is that it is non zero already at one loop. At one loop it contains a single excitation propagating freely through the flux tube. 

The next step is to decompose $r$ at one loop in terms of excitations propagating on the flux tube represented by the reference square. At one loop, the propagation of these excitations are organized in $SL(2)$ conformal blocks. At this point we face a technical obstacle. The $SL(2)$ conformal transformations form a three parameter group. The heptagon on the other hand has six independent conformal cross-ratios. In general all of them transform under the $SL(2)$ symmetry which preserves the two null lines associated to the channel under consideration. That makes the decomposition very hard.

Fortunately, there is a simple way around which --  in addition -- will also allow  us to make many more predictions at higher loops. Instead of choosing a single channel for the OPE expansion, we choose two adjacent channels. We then study excitations that propagate in both channels.
\begin{figure}[t]
\def\svgwidth{13cm}
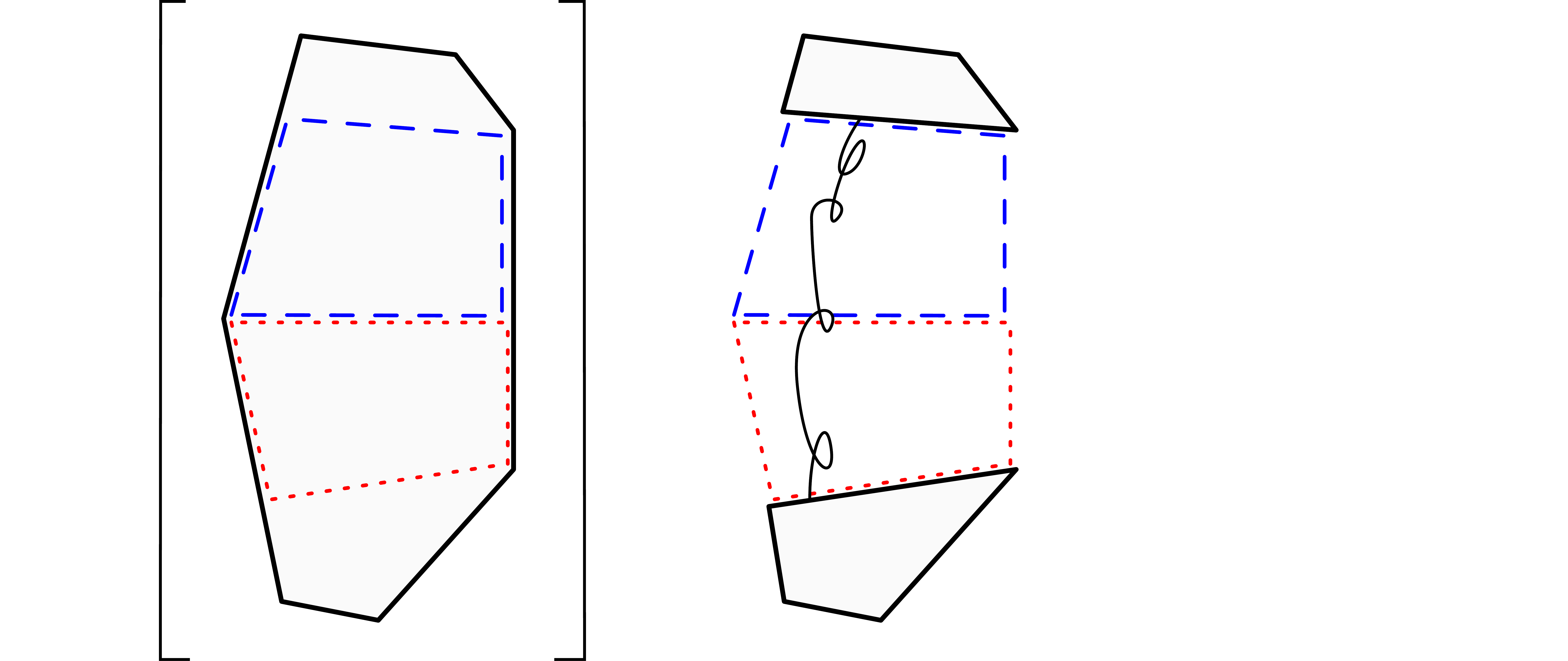
\caption{At one loop $\widetilde r$ is given by the correlation function of two Wilson loops as represented in the figure. The $u_i$ are given by 
$
u_{1,\dots,7}= \left\{\frac{x_{2,7}^2 x_{3,6}^2}{x_{2,6}^2 x_{3,7}^2},\frac{x_{1,3}^2
  x_{4,7}^2}{x_{1,4}^2 x_{3,7}^2},\frac{x_{1,5}^2 x_{2,4}^2}{x_{1,4}^2
  x_{2,5}^2},\frac{x_{2,6}^2 x_{3,5}^2}{x_{2,5}^2 x_{3,6}^2},\frac{x_{3,7}^2
  x_{4,6}^2}{x_{3,6}^2 x_{4,7}^2},\frac{x_{1,4}^2 x_{5,7}^2}{x_{1,5}^2
  x_{4,7}^2},\frac{x_{1,6}^2 x_{2,5}^2}{x_{1,5}^2 x_{2,6}^2}\right\}$; only six of them are independent.  }\label{tilder1loop}
\end{figure}
For the heptagon, any such choice is related by cyclic permutation. In figure \ref{TopBottom}b we plotted two reference squares, red (dotted) and blue (dashed), corresponding to the two adjacent channels. We will refer to these as the {\it bottom} (red dotted) and {\it top} (blue dashed) channels. Instead of the ratio $r$ (\ref{smallr}), we now consider the new finite conformal invariant ratio     
\beq\la{rpentagon}
\widetilde r=\log {\<W^{\text{heptagon}}\>\<W^\text{ref pentagon}\>\over\<W^\text{top hexagon}\>\<W^\text{bot hexagon}\>} \ . 
\eeq
That ratio is constructed such that it only contain excitations that propagate through both flux tubes represented by the red (dotted) and blue (dashed) squares. It is related to $r$ and to the remainder function in a simple way that only involves lower point objects that we know already, 
\beqa\la{remainder}
\widetilde r&=& r-\log {\<W^\text{bot hexagon}\>\<W^\text{red dotted square}\>\over\<W^\text{ref pentagon}\> \<W^\text{top pentagon}\>}\nn\\&=&R_{\text{heptagon}}+\widetilde r_{\widetilde U(1)}-R_\text{ref pentagon}+R_\text{top hexagon}+R_\text{bot hexagon} \,.
\eeqa
Note that for $\widetilde r$, there is no distinction between what we call top and bottom; it  possesses a symmetry that exchanges the two parts of the polygon. 
%As for $r$, the new ratio $\widetilde r$ is non zero already at one loop. 
At one loop, $\widetilde r$ is given by the correlation function of two Wilson loops as represented in figure \ref{tilder1loop}. For the bootstrap program, we have to decompose the one loop result in excitations that propagate in both channels.

Each of the channels is defined by a choice of two null lines and automatically defines a corresponding top and bottom group of cusps. Each channel comes with its three parameter group of $SL(2)$ transformations which preserve the corresponding pair of null lines. These transformations are then taken to act on a group of cusps on one of its two sides.  This is how we would measure what is flowing, see previous section and \cite{blocks2}. For example, we can choose to act with the $SL(2)_{\text{bottom}}$ transformations on the bottom cusps $2,3,4$ (with respect to the red dotted square) and with the $SL(2)_{\text{top}}$ transformations on the top cusps $5,6,7$ (with respect to the blue dashed square), see figure \ref{TopBottom}. Since the two $SL(2)$'s act on different cusps they clearly commute. In particular, 
\beq
\textit{The Casimirs of the two $SL(2)$'s commute. }\nn
 \eeq
 That is, we can decompose $\widetilde r$ into functions that are \textit{at the same time} conformal blocks with respect to $SL(2)_\text{bottom}$ and $SL(2)_\text{top}$. That fact is the main reason for studying $\widetilde r$ instead of $r$. Without using these new blocks there would be no practical advantage; decomposing the new ratio $\widetilde r$ would have been as hard as decomposing the original ratio $r$. We will refer to these blocks as {\it Multichannel blocks}. The Multichannel blocks solve at the same time the differential equations arising from both Casimirs. With the appropriate boundary conditions, compatible with the OPE expansion, the Multichannel blocks are fixed uniquely in this way. Note that the Multichannel blocks, that are functions of the conformal cross-ratios, do not factorize into a conformal block of $SL(2)_\text{bottom}$ times a conformal block of $SL(2)_\text{top}$. That is because the conformal cross-ratios are constructed from ratios of distances of cusps on the bottom and top parts of the heptagon. Generically, all of them transform under both $SL(2)$'s.\footnote{In other words, the two differential equations, derived from the two Casimirs, do not admit solutions of factorized form.
Instead, we have to solve  both equations at the same time and the solution will entangle the cross-ratios in a non-trivial way.}

To describe the Multichannel blocks we reconsider the two reference squares in greater detail. 
Each reference square gives us a three parameter family of polygons parametrized by the three symmetries of the square $R_\tau\times R_\sigma\times SO(2)_\phi$. For the heptagon, we have two sets of these variables $\{\tau_b,\sigma_b,\phi_b\}$ and $\{\tau_t,\sigma_t,\phi_t\}$. Together, they give us a basis of six independent cross ratios as needed for the heptagon. We see that a choice of successive OPE channels results in a natural basis for the conformal cross-ratios associated with these channels. A choice of these can be nicely represented using the $Y$ functions of \cite{Ypaper} as\footnote{A choice of momentum twistors yielding these cross-ratios is given in the appendix} 
\beq
\begin{array}{l}\widehat Y_{1,1}=\frac{\left\langle \lambda _5,\lambda _1,\lambda _2,\lambda
   _3\right\rangle  \left\langle \lambda _6,\lambda _7,\lambda _1,\lambda
   _2\right\rangle }{\left\langle \lambda _5,\lambda _6,\lambda _1,\lambda
   _2\right\rangle  \left\langle \lambda _7,\lambda _1,\lambda _2,\lambda
   _3\right\rangle }=e^{\phi_b-\sigma_b-\tau_b}\\
   \widehat Y_{1,2}= \frac{\left\langle \lambda _4,\lambda _5,\lambda
   _1,\lambda _2\right\rangle  \left\langle \lambda _5,\lambda _6,\lambda
   _7,\lambda _1\right\rangle }{\left\langle \lambda _4,\lambda _5,\lambda
   _6,\lambda _1\right\rangle  \left\langle \lambda _5,\lambda _7,\lambda
   _1,\lambda _2\right\rangle }=e^{-2\tau_b}\\
   \widehat Y_{1,3}=\frac{\left\langle \lambda _4,\lambda
   _5,\lambda _6,\lambda _2\right\rangle  \left\langle \lambda _5,\lambda
   _6,\lambda _7,\lambda _1\right\rangle }{\left\langle \lambda _4,\lambda
   _5,\lambda _6,\lambda _7\right\rangle  \left\langle \lambda _5,\lambda
   _6,\lambda _1,\lambda _2\right\rangle }=e^{-\phi_b-\sigma_b-\tau_b}\\
\end{array}\ ,\qquad
\begin{array}{l}\widehat Y_{2,1}=\frac{\left\langle \lambda _3,\lambda _4,\lambda _5,\lambda
   _2\right\rangle  \left\langle \lambda _4,\lambda _5,\lambda _6,\lambda
   _1\right\rangle }{\left\langle \lambda _3,\lambda _4,\lambda _5,\lambda
   _6\right\rangle  \left\langle \lambda _4,\lambda _5,\lambda _1,\lambda
   _2\right\rangle }=e^{-\phi_t-\sigma_t-\tau_t}\\
   \widehat Y_{2,2}= \frac{\left\langle \lambda _4,\lambda _5,\lambda
   _2,\lambda _3\right\rangle  \left\langle \lambda _5,\lambda _6,\lambda
   _1,\lambda _2\right\rangle }{\left\langle \lambda _4,\lambda _5,\lambda
   _6,\lambda _2\right\rangle  \left\langle \lambda _5,\lambda _1,\lambda
   _2,\lambda _3\right\rangle }=e^{-2\tau_t}\\
   \widehat Y_{2,3}=\frac{\left\langle \lambda _4,\lambda
   _1,\lambda _2,\lambda _3\right\rangle  \left\langle \lambda _5,\lambda
   _7,\lambda _1,\lambda _2\right\rangle }{\left\langle \lambda _4,\lambda
   _5,\lambda _1,\lambda _2\right\rangle  \left\langle \lambda _7,\lambda
   _1,\lambda _2,\lambda _3\right\rangle }=e^{\phi_t-\sigma_t-\tau_t}\\
\end{array}\la{Ys}
\eeq
The Multichannel conformal blocks are functions of the two $SL(2)$ Casimirs $C_i=\beta_i(\beta_i-1)$ as well as on the two momenta $k_i$ with respect to the $\sigma_i$ directions. Here $i=t,b$ stands for top or bottom. They are Appell hypergeometric functions of the second kind  $F_2$ \cite{tables}
\beqa\la{mega}
\cF_{\beta_b,\beta_t,k_b,k_t}(T_b,T_t) &=& {e^{i \pi  \left(\beta _b+\beta
  _t\right)}\over T_b^{2\beta_b} T_t^{2\beta_t}}\sum_{n_b=0}^\infty\sum_{n_t=0}^\infty \frac{ \(-T_b^{-2}\)^{n_b} \(-T_t^{-2}\)^{n_t} }{n_b!\, n_t!} \times\\ &&
  \frac{
  \Gamma
  \left(-\frac{ik_b}{2}+n_b+\beta _b\right)
  \Gamma \left(-\frac{ik_t}{2}+n_t+\beta
  _t\right) \Gamma
  \left(-\frac{ik_b}{2}-\frac{ik_t}{2}+n_b+n
  _t+\beta _b+\beta _t\right)}{ 
  \Gamma \left(n_b+2 \beta _b\right)
  \Gamma \left(n_t+2 \beta _t\right)} \nn
\eeqa
where $T_i=e^{\tau_i}$. These Multichannel blocks are obtained by solving the corresponding two differential equation for the two reference squares as explained in \cite{Hexagonpaper}. They are normalized such that they are symmetric under the exchange of the top and bottom and are only functions of the Casimirs. That is, they are invariant under $\beta_i\to(1-\beta_i)$. 

Note that the Multichannel blocks do not depend on $\phi_b$ and $\phi_t$ in a direct way. Their only dependence on the corresponding charges is through $\beta_b$ and $\beta_t$. To understand what values the $\beta$'s can take and what is their dependence on the two $SO(2)_\phi$ charges $m_b$ and $m_t$ we have to discuss what type of primary excitations are propagating on the flux tube at one loop. These excitations were studied in detail in \cite{Benjamin,Hexagonpaper}. For any $SO(2)$ charge $m$, there are two types of excitations with $\beta=m/2$ and $\beta=-m/2$ corresponding to the two possible polarizations of the gluon. The type of excitation/gluon polarization does not depend on the reference square. Hence, an excitation with $\beta=\pm m/2$ will have the same relation in any square. In our convention for the signs of $\phi_b$ and $\phi_t$, it means that an excitation with $\beta_b=\pm m_b/2$ will have $\beta_t=\mp m_t/2$ correspondingly.  Finally, note that the ratio $\widetilde r$ is parity invariant. Parity inverts the signs of $\phi_b$ and $\phi_t$. Therefore $\widetilde r$ depends on $e^{m_b\phi_b+m_t\phi_t}$ only through the combination $\cosh(m_b\phi_b+m_t\phi_t)$. Given that content of excitations, parity and the symmetry of $\widetilde r$ under exchanging top and bottom, we are led to the general form of $\widetilde r_{U(1)}$
\beq\la{form}
\widetilde r_{U(1)}=\int\! \frac{dk_b}{2\pi} \int \,\frac{dk_t}{2\pi}\,e^{-ik_b\sigma_b-ik_t\sigma_t}\sum_{m_b,m_t=1}^\infty\mathcal{A}_{m_b,m_t,k_b,k_t}(\phi_b,\phi_t)\,\cF_{{m_t\over2},{m_b\over2},k_t,k_b}(e^{\tau_t},e^{\tau_b})
\eeq
where
\beqa\la{decomposition}
\mathcal{A}_{m_b,m_t,k_b,k_t}(\phi_b,\phi_t)&=&  \delta_{m_t \ge 1, m_b \ge 2} \,  \cC_{m_t,m_b-2,k_t,k_b}\cosh\(m_t\phi_t+(m_b-2)\phi_b\)\\
&+& \delta_{m_t \ge 2, m_b \ge 1} \,\cC_{m_b,m_t-2,k_b,k_t}\cosh\((m_t-2)\phi_t+m_b\phi_b\)\nn\\
&+&\delta_{m_t \ge 2, m_b \ge 2} \,\cC_{2-m_b,m_t-2,k_b,k_t}\cosh\((m_t-2)\phi_t+(2-m_b)\phi_b\)\nn\\&+&  \cC_{m_t,-m_b,k_t,k_b}\cosh\(m_t\phi_t-m_b\phi_b\)  \nn
\eeqa
where $\delta_X$ is equal to $1$ if $X$ is true and $0$ otherwise.
In (\ref{form}), we have used the symmetry of the $SL(2)$ casimir $C=\beta(\beta-1)$ under the exchange of $\beta$ with $1-\beta$ to represent all representation with $\beta\ge0$.

The one loop function $\widetilde r_{U(1)}$ for the heptagon is given in figure \ref{tilder1loop}. It can be written indeed in the OPE form (\ref{form}). Needless to say, this is a very non-trivial check of all the statements so far. To decompose  $\widetilde r_{U(1)}$ by brute force and hence read the form factors $\cC_{m_1,m_2,k_1,k_2}$ is not an easy task simply because the function we want to compute is a complicated function of transcendentality degree two. However it turns out that we can construct some very useful box operators $\Box_b\equiv \partial_{\phi_b}^2-\partial_{\sigma_b}^2$ and $\Box_t\equiv \partial_{\phi_t}^2-\partial_{\sigma_t}^2$. When acting on $\widetilde r$, these operators simplify it dramatically.\footnote{This strategy was used in \cite{Hexagonpaper} as well to decompose the hexagon one loop result, see section 5.3. We see that the same kind of operators also simplify dramatically the heptagon result. This strongly indicates that there is something deep about them and that the Hexagon simplification was not a simple coincidence.} For example the action of $\Box_b$ on $\widetilde r_{U(1)}$ yields a rational function of the $Y$-functions! This is highly non-trivial; naively we would expect lots of logs to survive and the expression to be a huge mess. Furthermore, further acting on this rational function with $\Box_t$ simplifies the result even more; once again, a priori, we would expect the exact opposite. We do not have a physical understanding of why these box operators are so remarkable, it would be very interesting to figure it out. At the end of the day all we need to do is to expand a (not so horrendous) rational function. To undo the action of the box operators we simply insert some propagator like terms. Then we check the decomposition numerically to be sure we did not loose any zero mode. The conclusion turns out to be remarkably simple,  
\beq
\cC_{m_1,m_2,k_1,k_2}=-\frac{2 e^{\frac{1}{2} i \pi  \left(m
  _1+m _2\right)} \Gamma
  \left(\frac{i k_1}{2}+\frac{i
  k_2}{2}+\frac{m _1}{2}+\frac{m
  _2}{2}\right)}{\left(k_1-i m
  _1\right) \left(k_2+i m _2\right)
  \Gamma \left(-\frac{i k_1}{2}-\frac{i
  k_2}{2}+\frac{m _1}{2}+\frac{m
  _2}{2}+1\right)}\ . \la{theC}
\eeq
In (\ref{form}) $k_b$ is integrated slightly below the real axis and $k_t$ slightly below the integration contour of $k_b$, that is $k_b \in \mathbb{R}-i0$ and $k_t \in \mathbb{R}-2i0$.

Given the one loop decomposition of $\widetilde r$, it is straightforward to obtain a big part of the higher loop result. E.g., the two-loop remainder function can be split into two pieces as  
\beq
\widetilde r_{\text{2 loops}}= \tau_t  \[\widetilde r_{\text{2 loops}}\]_{\text{$\tau_t$}}+\[\widetilde r_{\text{2 loops}}\]_{\text{$\tau_t^0$}} \la{twohalves}
\eeq
The piece linear in $\tau_t$, also called the OPE discontinuity, is  obtained by dressing (\ref{form}) by the one loop anomalous dimension $\gamma_{2s_t}(k_t)$ that is  given by \cite{bootstraping, Benjamin}
\beq
\gamma_{2s}(k)=\gamma_{2-2s}(k)=2g^2\[\psi(s+i{k\over2})+\psi(s-i{k\over2})-\psi(1)\]
\eeq
Here, $s$ is the conformal spin of the excitation. For $\beta=|m|/2$ in (\ref{decomposition}) we have $s=1+\beta$ and for $\beta=1+|m|/2$ in (\ref{decomposition}), we have $s=|m|/2$. That is, $\[\widetilde r_{\text{2 loops}}\]_{\text{$\tau_t$}}$ is obtained from $\widetilde r_{U(1)}$ (\ref{form}) by dressing $\cA$ as
\beqa\la{twoloops}
\mathcal{A}_{m_b,m_t,k_b,k_t}(\phi_b,\phi_t)\quad\to&&  \gamma_{2+m_t}(k_t)\,\delta_{m_t \ge 1, m_b \ge 2} \,  \cC_{m_t,m_b-2,k_t,k_b}\cosh\(m_t\phi_t+(m_b-2)\phi_b\)\\
&+& \gamma_{m_t}(k_t)\quad \delta_{m_t \ge 2, m_b \ge 1} \,\cC_{m_b,m_t-2,k_b,k_t}\cosh\((m_t-2)\phi_t+m_b\phi_b\)\nn\\
&+& \gamma_{m_t}(k_t)\quad\delta_{m_t \ge 2, m_b \ge 2} \,\cC_{2-m_b,m_t-2,k_b,k_t}\cosh\((m_t-2)\phi_t+(2-m_b)\phi_b\)\nn\\&+&  \gamma_{2+m_t}(k_t)\, \cC_{m_t,-m_b,k_t,k_b}\cosh\(m_t\phi_t-m_b\phi_b\)  \nn
\eeqa
More generally, at $l$ loops there are $l$ terms we can predict.\footnote{It would be interesting to understand the Regge theory interpretation of these terms \cite{Bartels:2011xy}. There seems to be a very interesting connection between the OPE expansion and the BFKL approach \cite{BFKL} as explained  in \cite{Bartels:2011xy}.} These are the terms that are proportional to $\tau_t^{l-j}\tau_b^{j-1}$ with $j=1,2,...,l$. They are obtained from $\widetilde r$ by dressing the excitations with $\gamma_{2s_t}(k_t)^{l-j}\gamma_{2s_b}(k_b)^j$ as in (\ref{twoloops}). The corresponding terms in the remainder function are easy to read from (\ref{remainder}). 

These are $l$ new prediction at any loop order $l$. The next step involves a well posed mathematical problem. These predictions gives us part of a remainder function which we want to compute. We can repeat the computation using other OPE channels and hence obtain several constraints on the remainder function. The mathematical problem is to find a function with all the required symmetries and with the good OPE expansion in all possible channels. At two loops, if we assume that the answer has a symbol, one can show that the contributions computed above suffice to uniquely fix the function \cite{OPEpaper}. It would be interesting to see if these terms predicted here can fulfill the same role at higher loops. 

Finally note that our prediction expressed here as infinite sums should be equal to some function of transcendentality degree three. In \cite{Hexagonpaper} we succeed in resuming similar sums and got an explicit expression for the transcendental function.  For the heptagon, we did not succeed in doing so. On the other hand, the sums yield a perfectly fine representation of the corresponding pieces of the answer; in particular these sums converge very fast (exponentially fast for large $\tau_i$).

It is also possible to replace the decomposition mentioned above by the action of some convolution kernels and projection operators on the one loop result, see section 3.5 in  \cite{Hexagonpaper}. The two approaches are equivalent, they are basically the Fourier transform of each other. 

Finally let us conclude with a discussion of the relation of our findings with those of \cite{Simon}. In \cite{Simon} the symbols of all two loop MHV amplitudes were computed. It would be very interesting to extract the seven point MHV amplitude from its symbol and compare its OPE discontinuity against our predictions. Alternatively, it would be great to compute the symbol of our sum, use it to construct the full symbol following \cite{Hexagonpaper}, and compare it against the symbol derived in \cite{Simon}. These two projects involve considerable work.

We conclude with a third proposal which is considerably simpler to implement and which allows to "see" a great deal of the OPE expansion \textit{structure} from the symbol of \cite{Simon}. The proposal is to extract the symbol of the OPE discontinuity $\mathbb{D}_2$ from the symbol computed by Caron-Huot \cite{Simon} and check that it satisfies the equation $D_+D_- \mathbb{D}_2=0$. Here, $D_\pm=\cC-\partial_{\phi}\(\partial_{\phi}\pm 2\)$ projects out one of the two types of excitation, see \cite{Hexagonpaper} for more details.  A more restricted check can be done using the multichannel structure. For the heptagon for example, consider the symbol of the combination $\widetilde r$ (\ref{remainder}) and extract from it the symbol of the OPE discontinuity in either of the two channels. This is a simpler symbol $\widetilde{\mathbb D}$ with three slots.  Then, from the OPE decomposition and the relation (\ref{remainder}) it follows that 
\beq
D_+^{(b)}D_-^{(b)}\widetilde{\mathbb D}_2=D_+^{(t)}D_-^{(t)}\widetilde{\mathbb D}_2=D_-^{(b)}D_-^{(t)}\widetilde{\mathbb D}_2=D_+^{(t)}D_+^{(b)}\widetilde{\mathbb D}_2=0\nn
\eeq
where $D_{\pm}^{(i)}=\cC_i-\partial_{\phi_i}\(\partial_{\phi_i}\pm 2\)$ and the Casimir differential operators are written in the appendix.
%     
%very simple:
%\begin{itemize}
%\item First take the symbol as computed by Caron-Huot \cite{Simon} plus the symbol of the top and bottom Hexagons computed in \cite{Gon} (see also \cite{Hexagonpaper}). These are symbols with four slots.
%\item Combine them as in (\ref{remainder}). Extract the symbol of the OPE discontinuity from this symbol by identifying the first slot \cite{Hexagonpaper}. This is a simpler symbol $\mathbb{D}$ with three slots.
%\item Then, from the OPE picture it follows that 
%\beqa
%0&=&\[\mathcal{C}_b-\partial_{\phi_b}\(\partial_{\phi_b}+2\)\]\circ\[\mathcal{C}_b-\partial_{\phi_b}\(\partial_{\phi_b}-2\)\]\circ \mathbb{D} \nn \\
%0&=&\[\mathcal{C}_b-\partial_{\phi_t}\(\partial_{\phi_t}+2\)\]\circ\[\mathcal{C}_t-\partial_{\phi_t}\(\partial_{\phi_t}-2\)\]\circ \mathbb{D}\nn \\
%0&=&\[\mathcal{C}_b-\partial_{\phi_b}\(\partial_{\phi_b}+2\)\]\circ\[\mathcal{C}_t-\partial_{\phi_t}\(\partial_{\phi_t}+2\)\]\circ \mathbb{D}\nn\\
%0&=&\[\mathcal{C}_b-\partial_{\phi_b}\(\partial_{\phi_b}-2\)\]\circ\[\mathcal{C}_t-\partial_{\phi_t}\(\partial_{\phi_t}-2\)\]\circ \mathbb{D} \nn
%\eeqa
%where the Casimir differential operator $\cC$ is given in the appendix. These relations should hold as a simple consequence of the differential equations for the conformal blocks plus the structure of the expansion (\ref{form}). These conditions generalize the equation $D_+D_- D_2=0$ used in \cite{Hexagonpaper}, see equation (83) there. 
%\end{itemize}
This third proposal has a clear advantage over the other two: it is straightforward to implement. Acting with differential operators on symbols is very simple, after all one of the definitions of symbols is through its differential, see e.g. \cite{Gon}. The disadvantage of this check is that it does not probe the full OPE expansion. Namely, it probes the structure (\ref{form}),(\ref{twoloops}) but it is insensitive to the form of the form factors. To probe the precise form of the form factors we have to work harder as described above. 

\section{Generalization to more sides}
\begin{figure}[t]
\centering
\def\svgwidth{14cm}
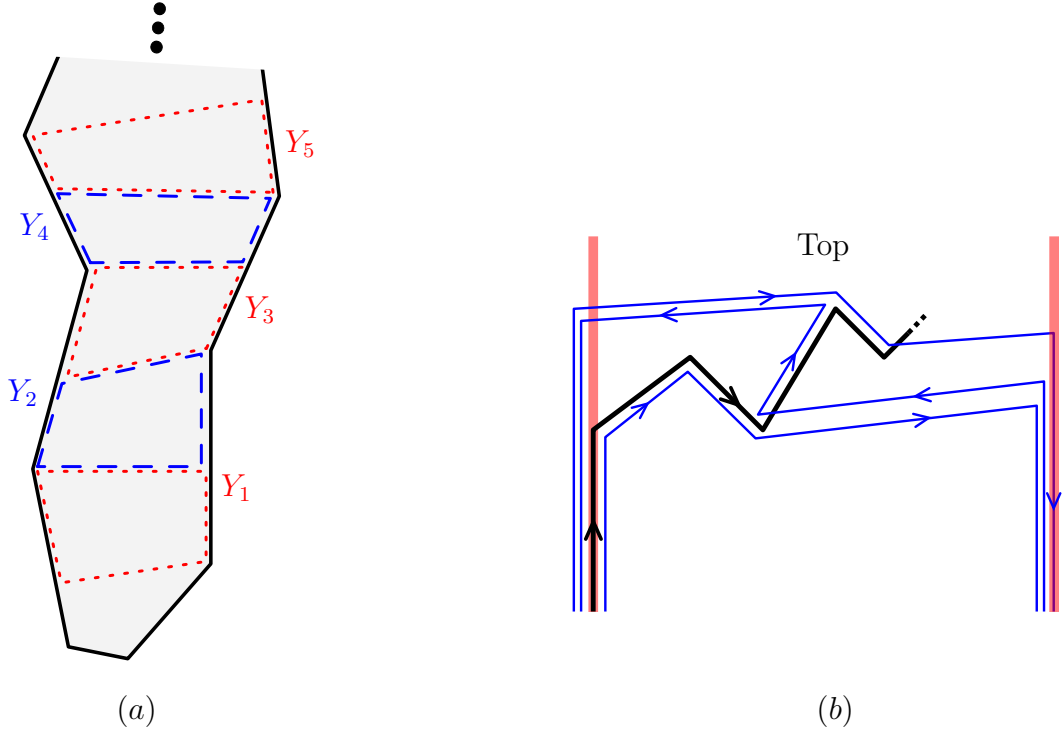
\caption{(a) A particularly useful choice of successive channels convenient for describing NPWL using multichannel blocks. (b)Any null polygon Wilson loop can be decomposed into a sum of octagons, heptagons and hexagons, all sharing two null lines of the OPE expansion. The figure demonstrate such a decomposition of the top part of the loop (black) in these building blocks (blue thin lines). The two thick red lines represent the null lines of the OPE expansion.
}\label{Linear}
\end{figure}

\la{sec3}
The techniques applied in the previous section to the heptagon generalize in a straightforward way to polygons with more edges. For an $n$ sided polygon, one choses a sequence of $n-5$ successive channels as done in figure \ref{TopBottom}.b for the heptagon. That is, let the first channel on the "bottom" be associated with the two null edges $(j,j+3)$. Then the next channel be associated with the two null edges $(j,j+4)$ or $(j-1,j+3)$ and so on. For a given bottom channel, there are $2^{n-6}$ possible choices for the successive channels. For example, one choice is to have all the channels sharing the edge $j$. A natural symmetric choice is given by the $Y$ function of \cite{Ypaper}. That is the choice where the channels alternate as $(j,j+3)\to(j,j+4)\to(j+1,j+4)\to(j+1,j+5)\to\dots$, see figure \ref{Linear}.a. Each of these channels comes with its $SL(2)$ conformal group. The actions of all these $SL(2)$'s commute.

Next, for each channel one choose an associated reference square. A natural choice that generalize our choice for the heptagon is again in terms of the $\widehat Y$ functions (see appendix D.2 in \cite{Ypaper})
\beq
\widehat Y_{k,1}=e^{\phi_k-\sigma_k-\tau_k}\ ,\qquad
   \widehat Y_{k,2}= e^{-2\tau_k}\ ,\qquad
   \widehat Y_{k,3}=e^{-\phi_k-\sigma_k-\tau_k}\nn
\eeq
Here, $\{\tau_i,\sigma_i,\phi_i\}$ parametrize the three isometries of the $i$'th square acting on the corresponding bottom group of cusps. The corresponding $\widetilde r$ for such sequence of squares is defined as in (\ref{rpentagon}) using a reference $n-2$ polygon. It is trivially related to the remainder function and lower $n$ polygons as in (\ref{remainder}). 

The Multichannel conformal blocks corresponding to the sequence of OPE channels are now functions of $\{\beta_i,k_i,\tau_i\}$ where $i=1,\dots,n-5$ parametrize the channel. They are obtained by solving $n-5$ differential equations and are expected to be generalized hypergeometric functions of $n-5$ variables. A decomposition of $\widetilde r$ in terms of the Multichannel block now allows for many predictions at higher loops.    

Given all the above, there is however a simpler way of decomposing the one loop result $r_{U(1)}$ for polygon with more then eight sides. It is based on the linearity of the one loop result and was first used in \cite{bootstraping} to compute the two-loop remainder function for all polygons in $\mathbb R^{1,1}$ kinematics. The idea is to decompose any higher $n$ polygon into a linear combination of octagons, heptagons and hexagons, all sharing two of their edges on the two null lines of the OPE expansion. That can be done as illustrated in figure \ref{Linear}.b. Due to the linearity on the $U(1)$ result, evaluating $r_{U(1)}$ on the high $n$ polygon is equal to the sum of the $U(1)$ result on these smaller building blocks. In that way, a decomposition of the octagon, heptagon and hexagon automatically result in a decomposition of any higher sided polygon. 

Another way of getting the decomposition of $r_{U(1)}$ for any polygon is by using the triangles of \cite{Hexagonpaper}. That is, $r_{U(1)}$ is the one loop correlation function of two polygons as illustrated in figures \ref{roctagon}.b and \ref{tilder1loop}. As explained in \cite{Hexagonpaper}, each of these two polygons can be decomposed as a sum of triangles. Therefore, the one loop correlator decomposed into a sum of correlators of pairs of triangles, one from the top and the other from the bottom. A decomposition of the contribution of such pair of triangle into excitations propagating in the flux tube automatically gives a decomposition of $r_{U(1)}$ for any polygon. 

Both of these alternative approaches lead to a decomposition of the $U(1)$ result in a single channel. They provide us with a restricted set of prediction for higher loops compared to the multichannel decomposition, see previous section.

Finally let us end with a few words about the ground-breaking recent work done by S. Caron-Huot which appeared recently in \cite{Simon}. In this paper the symbol of all MHV amplitudes at two loops was conjectured.  We hope that the predictions presented in this paper for the remainder function will be useful to cross-check the results of \cite{Simon} and in particular to help construct the Heptagon two-loop remainder function from its symbol. This was our main motivation for writing this short note. In the longer term we hope that the higher loop predictions presented in this paper will be useful in understanding the all loops result.

\section*{Acknowledgments}

We thank D.\ Gaiotto, J.\ Maldacena, J.\ Penedones, D.\ Skinner, T.\ Wang,  the participants of the \textit{The Harmony of Scattering Amplitudes} program at KITP for  useful discussions. We are grateful to F.\ Alday, D.\ Gaiotto, and J.\ Maldacena for collaboration on closely related topics. We are specially grateful to J.\ Maldacena's  and J.\ Penedones for  comments on the manuscript and for very enlightening discussions.  
The research of A.S. and P.V. has been supported in part by the Province of Ontario through ERA grant ER 06-02-293. Research at the Perimeter Institute is supported in part by the Government of Canada through NSERC and by the Province of Ontario through MRI. A.S. and
P.V. would like to thank KITP for warm hospitality during the period of completion of this work.
This work was partially funded by the research grants PTDC/FIS/099293/2008 and CERN/FP/116358/2010. This research was supported in part by the National Science Foundation under Grant No. NSF PHY05-51164.

\appendix
\section{Heptagon details}
In this appendix we present some details concerning the heptagon computation. It is often useful to have a parametrization of the polygons in terms of momentum twistors. A particularly simple such choice for the heptagon which leads to the cross ratios (\ref{Ys}) is
\beqa
&&\lambda_1=\(1,0,0,0\) \, , \qquad \lambda_2=\(-1,0,0,1\) \,,\qquad  \lambda_4=\(0,1,-1,1\) \,, \qquad \lambda_5=\(0,1,0,0\)\nn \\
&& \lambda_3=\(-1 , (e^{-\tau_t+\sigma_t+\phi_t}+e^{-2\tau_t+2\phi_t})^{-1},-e^{\tau_t-\sigma_t-\phi_t},e^{-\tau_t-\sigma_t-\phi_t}+e^{\tau_t-\sigma_t-\phi_t}+1\) \,, \nn\\
&&\lambda_6=\(0,1,e^{\tau_b+\sigma_b-\phi_b}+e^{\phi_t-\phi_b+\sigma_b+\tau_b-\tau_t-\sigma_t},0\) \,,\qquad 
 \lambda_7=\(1,0,e^{-\phi_b-\sigma_b+\tau_b},e^{-\phi_b-\sigma_b-\tau_b}\)\nn
\eeqa
Similarly, to compute the ratio (\ref{rpentagon}) it is also useful to have a list of momentum twistors for the pentagon and for both hexagons, 
\beqa
\text{pentagon}&=& \{\lambda_1,\lambda_2,\lambda_{3+1/2},\lambda_5,\lambda_{6+1/2}\}\nn\\
\text{top hexagon}&=& \{\lambda_1,\lambda_2,\lambda_{3},\lambda_4,\lambda_5,\lambda_{6+1/2}\}\nn \\
\text{bot hexagon}&=& \{\lambda_1,\lambda_2,\lambda_{3+1/2},\lambda_5,\lambda_6,\lambda_7\}\nn
\eeqa
where 
\beq
\lambda_{3+1/2}=\(0,(1+e^{\phi_t-\tau_t-\sigma_t})^{-1},-1,1\) \, \qquad \lambda_{6+1/2}=\(0,0,1,0\) \,.\nn
\eeq
There are two more polygons which appear in our discussion: the top and bottom reference squares. The mometum twistors of the bottom square can be parametrized as
\beq
\{\lambda_\text{left},\lambda_\text{top},\lambda_\text{right},\lambda_\text{bottom}\}_{\text{bottom}}=\{ \lambda_1 ,(0,0,0,1) ,\lambda_5 , \lambda_{6+1/2} \}\la{square}
\eeq
The Multichannel blocks are derived by acting with the two $SL(2)$ Casimirs discussed in the text. Each $SL(2)$ is the symmetry group which preserves two null lines. We have two $SL(2)$ because we use two pairs of null lines, one pair for each of the reference squares in figure \ref{TopBottom}b. The action of each of the Casimirs is derived exactly as in \cite{Hexagonpaper}, see section 5.2. Hence let us simply quote the final result. The differential equation for the bottom square reads
\beq\la{diff}
\Big(\mathcal{C}_b -4 \beta_b(\beta_b-1) \Big)e^{-ik_b \sigma_b -i k_t \sigma_t}\cF_{\beta_b,\beta_t,k_b,k_t}(T_b,T_t) =0
\eeq
with 
\beq
\mathcal{C}_b=e^{-2\tau_b} \[ \partial_{\tau_t} \partial_{\tau_b}- \partial_{\tau_t} \partial_{\sigma_b}  -\partial_{\sigma_t} \partial_{\tau_b}+ \partial_{\sigma_t} \partial_{\sigma_b}+(1+e^{2\tau_b}) \partial^2_{\tau_b} -2 \partial_{\tau_b}\partial_{\sigma_b}+\partial^2_{\sigma_b}+2 e^{2\tau_b} \partial_{\tau_b}\] \nn
\eeq
The differential equation for the top channel is obtain from (\ref{diff}) with $\mathcal{C}_b\to \mathcal{C}_t$ and $\beta_b\to \beta_t$. The two differential equations  define the Multichannel block (\ref{mega}). In the two dimensional octagon case the Casimir operator was much simpler; it was simply given by (\ref{casimiroctagon}). 

Concerning the decomposition of the $1$ loop result, we find the remarkably simple result
\beq
\Box_t \Box_b \widetilde r_{U(1)} = -\frac{32 Y_{11} Y_{12}^2 Y_{13} Y_{21} Y_{22}^2 Y_{23}}{\left(Y_{11} Y_{13}
   Y_{22}+Y_{12} \left(Y_{21} Y_{23}+\left(Y_{13}+Y_{21}+1\right) Y_{22}
   \left(Y_{11}+Y_{23}+1\right)\right)\right){}^3} \,P   \la{remarkable}
   \eeq
   where
\beqa
P&=&Y_{11} \left(Y_{12}+1\right) \left(\left(Y_{13}+1\right) Y_{22}+Y_{21}
   \left(Y_{22}+1\right)\right)+Y_{13} \left(Y_{23}+Y_{22}
   \left(Y_{23}+1\right)\right) \nn \\&+&Y_{12} \left(Y_{21} \left(Y_{22}+1\right)
   \left(Y_{23}+1\right)+\left(Y_{13}+1\right) \left(Y_{23}+Y_{22}
   \left(Y_{23}+1\right)\right)\right) \,. \nn
\eeqa
It is probably possible to simply this expression even further using an overcomplete set of $Y$-functions related to the basis used in this paper by the Y-system relations \cite{Ypaper}. 
Expression (\ref{remarkable}) is much simpler to decompose than $ \widetilde r_{U(1)} $ given in figure \ref{tilder1loop}. At the end, the action of the box operators is then  simple to undo in Fourier by inserting some propagator like factors. In this way one derives (\ref{theC}).


\begin{thebibliography}{99}

\bibitem{AmplitudeWilson}
  L.~F.~Alday, J.~M.~Maldacena,
  ``Gluon scattering amplitudes at strong coupling,''
  JHEP {\bf 0706}, 064 (2007).
  [arXiv:0705.0303 [hep-th]].
  $\bullet$  G.~P.~Korchemsky, J.~M.~Drummond, E.~Sokatchev,
  ``Conformal properties of four-gluon planar amplitudes and Wilson loops,''
  Nucl.\ Phys.\  {\bf B795}, 385-408 (2008).
  [arXiv:0707.0243 [hep-th]].
$\bullet$  A.~Brandhuber, P.~Heslop, G.~Travaglini,
  ``MHV amplitudes in N=4 super Yang-Mills and Wilson loops,''
  Nucl.\ Phys.\  {\bf B794}, 231-243 (2008).
  [arXiv:0707.1153 [hep-th]].
  $\bullet$  Z.~Bern, L.~J.~Dixon, D.~A.~Kosower, R.~Roiban, M.~Spradlin, C.~Vergu and A.~Volovich,
  ``The Two-Loop Six-Gluon MHV Amplitude in Maximally Supersymmetric Yang-Mills
  Theory,''
  Phys.\ Rev.\  D {\bf 78}, 045007 (2008)
  [arXiv:0803.1465 [hep-th]].
$\bullet$  J.~M.~Drummond, J.~Henn, G.~P.~Korchemsky and E.~Sokatchev,
  ``Hexagon Wilson loop = six-gluon MHV amplitude,''
  Nucl.\ Phys.\  B {\bf 815} (2009) 142
  [arXiv:0803.1466 [hep-th]].
$\bullet$  N.~Berkovits, J.~Maldacena,
  ``Fermionic T-Duality, Dual Superconformal Symmetry, and the Amplitude/Wilson Loop Connection,''
  JHEP {\bf 0809}, 062 (2008).
  [arXiv:0807.3196 [hep-th]].
$\bullet$   L.~J.~Mason, D.~Skinner,
  ``The Complete Planar S-matrix of N=4 SYM as a Wilson Loop in Twistor Space,''
  JHEP {\bf 1012}, 018 (2010).
  [arXiv:1009.2225 [hep-th]]. $\bullet$ S.~Caron-Huot,
  ``Notes on the scattering amplitude / Wilson loop duality,''
 [arXiv:1010.1167 [hep-th]]. $\bullet$   M.~Bullimore and D.~Skinner,
  ``Holomorphic Linking, Loop Equations and Scattering Amplitudes in Twistor Space,''
  arXiv:1101.1329 [hep-th]. $\bullet$   A.~V.~Belitsky, G.~P.~Korchemsky and E.~Sokatchev,
  ``Are scattering amplitudes dual to super Wilson loops?,''
  arXiv:1103.3008 [hep-th].
  
%\cite{Alday:2010ku}
\bibitem{OPEpaper}
  L.~F.~Alday, D.~Gaiotto, J.~Maldacena, A. Sever and P. Vieira,
  ``An Operator Product Expansion for Polygonal null Wilson Loops,''
 [arXiv:1006.2788 [hep-th]].

\bibitem{blocks1}
  F.~A.~Dolan and H.~Osborn,
  ``Conformal four point functions and the operator product expansion,''
  Nucl.\ Phys.\  B {\bf 599} (2001) 459
  [arXiv:hep-th/0011040].   
  %\cite{Dolan:2003hv}
  
\bibitem{blocks2}
  F.~A.~Dolan and H.~Osborn,
  ``Conformal partial waves and the operator product expansion,''
  Nucl.\ Phys.\  B {\bf 678} (2004) 491
  [arXiv:hep-th/0309180].
  %%CITATION = NUPHA,B678,491;%%
  
%\cite{Gaiotto:2011dt}
\bibitem{Hexagonpaper}
  D.~Gaiotto, J.~Maldacena, A.~Sever and P.~Vieira,
  ``Pulling the straps of polygons,''
  arXiv:1102.0062 [hep-th].
  %%CITATION = ARXIV:1102.0062;%%

%\cite{DelDuca:2009au}
\bibitem{DelDuca}
  V.~Del Duca, C.~Duhr and V.~A.~Smirnov,
  ``An Analytic Result for the Two-Loop Hexagon Wilson Loop in N = 4 SYM,''
  JHEP {\bf 1003} (2010) 099
  [arXiv:0911.5332 [hep-ph]].
  %%CITATION = JHEPA,1003,099;%%

  %\cite{Goncharov:2010jf}
\bibitem{Gon}
  A.~B.~Goncharov, M.~Spradlin, C.~Vergu and A.~Volovich,
  ``Classical Polylogarithms for Amplitudes and Wilson Loops,''
  Phys.\ Rev.\ Lett.\  {\bf 105}, 151605 (2010)
  [arXiv:1006.5703 [hep-th]].
  %%CITATION = PRLTA,105,151605;%%

%\cite{Gaiotto:2010fk}
\bibitem{bootstraping}
D.~Gaiotto, J.~Maldacena, A.~Sever and P.~Vieira,
  ``Bootstrapping Null Polygon Wilson Loops,''
  arXiv:1010.5009 [hep-th].

%\cite{Alday:2009yn}
\bibitem{AMapril}
  L.~F.~Alday and J.~Maldacena,
  ``Null polygonal Wilson loops and minimal surfaces in Anti-de-Sitter space,''
  JHEP {\bf 0911}, 082 (2009)
  [arXiv:0904.0663 [hep-th]].
  %%CITATION = JHEPA,0911,082;%%


%\cite{Drummond:2007aua}
\bibitem{Drummond:2007aua}
  J.~M.~Drummond, J.~Henn, G.~P.~Korchemsky and E.~Sokatchev,
  ``Conformal Ward identities for Wilson loops and a test of the duality with
  gluon amplitudes,''
  Nucl.\ Phys.\  B {\bf 826} (2010) 337
  [arXiv:0712.1223 [hep-th]].
  %%CITATION = NUPHA,B826,337;%%  %%CITATION = NUPHA,B795,385;%%

\bibitem{Benjamin}
  B.~Basso,
  ``Exciting the GKP string at any coupling,''
  arXiv:1010.5237 [hep-th].
  %%CITATION = ARXIV:1010.5237;%%
  
  %\cite{DelDuca:2010zp}
\bibitem{DelDucaOct}
  V.~Del Duca, C.~Duhr and V.~A.~Smirnov,
  ``A Two-Loop Octagon Wilson Loop in N = 4 SYM,''
  JHEP {\bf 1009} (2010) 015
  [arXiv:1006.4127 [hep-th]].
  
%\cite{Heslop:2010kq}
\bibitem{Heslop:2010kq}
  P.~Heslop, V.~V.~Khoze,
  ``Analytic Results for MHV Wilson Loops,''
  JHEP {\bf 1011}, 035 (2010).
  [arXiv:1007.1805 [hep-th]].
  
\bibitem{Simon}
S.~Caron-Huot, 
  ``Superconformal symmetry and two-loop amplitudes in 
planar N=4 super Yang-Mills''.

%\cite{Bern:2005iz}
\bibitem{BDS}
  Z.~Bern, L.~J.~Dixon and V.~A.~Smirnov,
  ``Iteration of planar amplitudes in maximally supersymmetric Yang-Mills
  theory at three loops and beyond,''
  Phys.\ Rev.\  D {\bf 72} (2005) 085001
  [arXiv:hep-th/0505205].
  %%CITATION = PHRVA,D72,085001;%%

\bibitem{Ypaper}
  L.~F.~Alday, J.~Maldacena, A.~Sever and P.~Vieira,
  ``Y-system for Scattering Amplitudes,''
  J.\ Phys.\ A  {\bf 43}, 485401 (2010)
  [arXiv:1002.2459 [hep-th]].
  %%CITATION = JPAGB,A43,485401;%%
  
  
\bibitem{tables}
J.~ Murley, N.~ Saad, 
"Tables of the Appell Hypergeometric Functions $F_2$", [arXiv:0809.5203 [math-ph]] $\bullet$ V.~V.~Bytev, M.~Y.~Kalmykov, B.~A.~Kniehl, "HYPERDIRE: HYPERgeometric functions DIfferential REduction MATHEMATICA based packages for differential reduction of generalized hypergeometric functions: now with pFq, F1,F2,F3,F4", [arXiv:1105.3565 [math-ph]]

\bibitem{BFKL}
  L.~N.~Lipatov,
  ``Reggeization of the Vector Meson and the Vacuum Singularity in Nonabelian
  Gauge Theories,''
  Sov.\ J.\ Nucl.\ Phys.\  {\bf 23} (1976) 338
  [Yad.\ Fiz.\  {\bf 23} (1976) 642]. $\bullet$   V.~S.~Fadin, E.~A.~Kuraev and L.~N.~Lipatov,
  ``On the Pomeranchuk Singularity in Asymptotically Free Theories,''
  Phys.\ Lett.\  B {\bf 60} (1975) 50. $\bullet$   E.~A.~Kuraev, L.~N.~Lipatov and V.~S.~Fadin,
  ``Multi - Reggeon Processes in the Yang-Mills Theory,''
  Sov.\ Phys.\ JETP {\bf 44} (1976) 443
  [Zh.\ Eksp.\ Teor.\ Fiz.\  {\bf 71} (1976) 840]. $\bullet$   I.~I.~Balitsky and L.~N.~Lipatov,
  ``The Pomeranchuk Singularity in Quantum Chromodynamics,''
  Sov.\ J.\ Nucl.\ Phys.\  {\bf 28} (1978) 822
  [Yad.\ Fiz.\  {\bf 28} (1978) 1597].
  

%\cite{Bartels:2011xy}
\bibitem{Bartels:2011xy}
  J.~Bartels, L.~N.~Lipatov and A.~Prygarin,
  ``Collinear and Regge behavior of 2 -> 4 MHV amplitude in N = 4 super
  Yang-Mills theory,''
  arXiv:1104.4709 [hep-th].
  %%CITATION = ARXIV:1104.4709;%%


  

  
  
  


\end{thebibliography}
\end{document}